# Molecular Cloud Formation Via Thermal Instability of Finite Resistive Viscous Radiating Plasma with Finite Larmor Radius Corrections

**Sachin Kaothekar[1,*]**
[1]Department of Physics, Mahakal Institute of Technology, Ujjain- 456664 (M.P.), India.

[*]Corresponding authors e-mail: sac_kaothekar@rediff.com, sackaothekar@gmail.com,

The effect of radiative heat-loss function and finite ion Larmor radius (FLR) corrections on the thermal instability of infinite homogeneous viscous plasma has been investigated incorporating the effects of thermal conductivity and finite electrical resistivity for the formation of a molecular cloud. The general dispersion relation is derived using the normal mode analysis method with the help of relevant linearized perturbation equations of the problem. Furthermore the wave propagation along and perpendicular to the direction of external magnetic field has been discussed. Stability of the medium is discussed by applying Routh Hurwitz's criterion and it is found that thermal instability criterion determines the stability of the medium. We find that the presence of radiative heat-loss function and thermal conductivity modify the fundamental criterion of thermal instability into radiatively driven thermal instability criterion. In longitudinal direction FLR corrections, viscosity, magnetic field and finite resistivity have no effect on thermal instability criterion. The presence of radiative heat-loss function and thermal conductivity modify the fundamental thermal instability criterion into radiatively driven thermal instability criterion. Also the FLR corrections modify the growth rate of the Alfven mode. For transverse wave propagation FLR corrections, radiative heat-loss function, magnetic field and thermal conductivity modify the thermal instability criterion. From the curves it is clear that heat-loss function, FLR corrections and viscosity have stabilizing effect, while finite resistivity has destabilizing effect on the thermal modes. Our results show that the FLR corrections and radiative heat-loss functions affect the evolution of interstellar molecular clouds and star formation.

## 1. Introduction

The problem of thermal instability is widely investigated due to its relevance to the fragmentation of interstellar medium and its role in molecular cloud formation. Also, thermal instability of molecular clouds is connected to the cloud collapse and star formation. In addition to this it is well known that thermal and radiative effects do play an important role in the stability investigations. Several authors investigated the phenomenon of thermal instability arising due to heat-loss mechanism in plasma. The thermal instability arising due to various heat-loss mechanisms definitely is the cause of astrophysical condensation and the formation of large and small objects. In a medium with a positive heat derivative with respect to density, if an isobaric perturbation grows in time the thermal instability occurs and the emission rate of energy decreases. This process is thought to be possible in a number of astrophysical situations such as the gas in clusters of the galaxies, in the solar corona and in the interstellar medium. The thing which is less clear is the relative importance of this process in various circumstances. Thermal instability has many applications in astrophysical situations (e.g. a clumpy interstellar medium, stellar atmosphere, star formation, globular clusters and galaxy formation and many more situations). Field (1965) discussed the importance of thermal instability in the formation of solar prominences, condensation in planetary nebula and condensation of galaxies from the intergalactic medium. The instability may be driven by radiative cooling of optically thin gas system or by exothermic nuclear reactions. Hunter (1966) investigated the effect of density stratification on thermal instability in a magnetized isothermal atmosphere. Aggarwal & Talwar (1969) discussed magneto-thermal instability in a rotating gravitating fluid. Ibanez (1985) studied the sound and thermal waves in a fluid with an arbitrary heat-loss function. Bora & Talwar (1993) investigated the magneto-thermal instability with finite electrical resistivity and Hall current. Kim & Narayan (2003) discussed the thermal instability in clusters of galaxies with conduction taking the effect of radiative heat loss function. Stothers (2003) carried out the problem of radiative instability in stellar envelops. Radwan (2004) studied the self-gravitating instability of radiating rotating gas cloud streams with non-uniform velocity. Menou et al. (2004) gave the importance of radiative effects in the Sun's upper radiative zone. Inutsuka et al. (2005) studied the propagation of shock wave into warm neutral medium taking into account radiative heating and cooling, thermal conduction and viscosity. Shadmehri & Dib (2009) discussed the thermal instability in magnetized partially ionized plasma with charged dust particles and radiative cooling function. Shaikh et al. (2008) discussed the Jeans instability of thermally conducting plasma in a variable magnetic field with Hall current, finite conductivity and viscosity. Recently Kaothekar & Chhajlani (2010) examined self-gravitational instability of partially ionized plasma with radiative effects. More recently Kaothekar et al. (2012) investigated the effect of neutral collision and radiative heat-loss function on self-gravitational instability of viscous thermally conducting partially-ionized plasma.

[*]First and Corresponding authors e-mail: sackaothekar@gmail.com

Along with this in above discussed problems the effect of finite ion Larmor radius is not considered. In many astrophysical situations such as in interstellar and interplanetary plasmas the zero Larmor radius approximation is not valid. Several authors (Roberts & Taylor 1962, Jeffery & Taniuti 1966, Jukes 1964, Vandakurov 1964, and Sharma 1974) pointed out the importance of finite ion Larmor radius (FLR) effects in the form of magnetic resistivity, on the plasma instability. Herrnegger (1972) studied the effects of collision and gyroviscosity on thermal instability in a two-component plasma and concluded that the critical wave number becomes smaller with increasing gyroviscosity for finite Alfven numbers and showed that Jeans criterion is changed by FLR for wave propagating perpendicular to magnetic field. Bhatia & Chhonkar (1985) investigated the stabilizing effect of FLR on the instability of a rotating layer of self-gravitating plasma. Vaghela & Chhajlani (1989) investigated the stabilizing effect of FLR on magneto-thermal stability of resistive plasma through porous medium with thermal conduction. Islam & Balbus (2005) examined dynamics of the magnetoviscous instability. Ferraro (2007) showed the stabilizing effect of FLR on magneto-rotational instability. Marcu & Ballai (2007) showed the stabilizing effect of FLR on thermosolutal stability of two-component rotating plasma. Sandberg et al. (2007) investigated the stabilizing effect of FLR on the coupled trapped electron and ion temperature gradient modes. Devlen & Pekunlu (2010) studied the effect of FLR on weakly magnetized, dilute plasma. Devlen (2011) investigated the problem of anisotropic transport effects on dilute plasmas. Recently Kaothekar & Chhajlani (2014) carried out the problem of Jeans instability of self-gravitating rotating radiative plasma with FLR corrections. More recently Kaothekar et al. (2016) investigated Jeans instability of partially-ionized self-gravitating viscous plasma with Hall effect FLR corrections and porosity. Thus FLR effect is an important factor in discussion of thermal instability and other hydrodynamic instabilities.

In this connection the effect of FLR corrections which is one of the anisotropic transport effects is important in dilute partially ionized astrophysical plasmas. Prakash & Manchanda (1974) investigated the effects of FLR and Hall currents on thermosolutal instability of a partially ionized plasma in porous medium. Mehta and Bhatia (1988) examined the Larmor radius effects on the instability of a stratified partially-ionized plasma in the presence of Hall currents and magnetic resistivity. Sharma & Chhajlani (1998) carried out the problem of effect of FLR on the Rayleigh-Taylor instability of two component magnetized rotating plasma. Alperovich & Chaikovsky (1999) examined the effective conductivity of the magnetized bounded partially ionized plasma with random irregularities. Pandey & Wardel (2006) carried out the problem of ion dynamics and the magnetorotational instability in weakly ionized discs. Sandoval-Villalbazo & Garcia-Perciante (2007) investigated gravitational instability of a dilute fully ionized gas in the presence of the Dufour effect. Kunz (2008) carried out the problem of linear stability of weakly ionized, magnetized planar shear flows. Devlen (2011) investigated the anisotropic transport effects in dilute plasmas. Recently Jain & Sharma (2016) carried out the problem of Jeans instability of a rotating partially ionized and strongly coupled plasma with Hall current. So we conclude that FLR corrections is one of the anisotropic transport effects and is important in dilute astrophysical plasmas such as in the case of molecular cloud formation.

In the light of above work, we find that Bora & Talwar (1993) considered the effect of finite electrical resistivity, electron inertia, Hall current, thermal conductivity and radiative heat-loss function, but they neglect the effect of FLR corrections and viscosity on radiative instability. Vaghela & Chhajlani (1989) considered the effect of finite electrical resistivity, viscosity and thermal conductivity, but they neglect the effect of radiative heat-loss function on thermal instability. Aggarwal & Talwar (1969) considered the effect of viscosity, rotation, finite electrical resistivity, thermal conductivity and radiative heat-loss function, but they neglect the effect of FLR correction on radiative instability. Shaikh et al. (2008) considered the effect of viscosity, finite electrical resistivity, Hall current and thermal conductivity, but they neglect the effect of FLR corrections and radiative heat-loss function on thermal instability. Thus we find that in these studies (Vaghela & Chhajlani 1989, Shaikh et al. 2008, Aggarwal & Talwar 1969, Bora & Talwar 1993, and Kaothekar et al. 2016) the joint influence of FLR correction, radiative heat-loss function, viscosity, electrical resistivity, thermal conductivity and magnetic field on the thermal instability is not investigated. We also find that none of the above authors has tried to explore thermal instability criterion for viscous and non-viscous plasma with FLR corrections, radiative heat-loss function and thermal conductivity. Therefore in the present work thermal instability of magnetized plasma with FLR corrections, radiative heat-loss function, viscosity, thermal conductivity and finite electrical resistivity is studied. We also want to explore the importance of viscous and non-viscous system and its impact on the thermal instability of plasma in connection with FLR correction radiative heat-loss function, thermal conductivity and finite electrical resistivity. This work is applicable to interstellar molecular clouds formation and formation of stars.

This paper is organized as follows. Section 2 contains the basic equations for a magneto thermal system. In section 3 linearized equations and the dispersion relation are derived for the first order approximation. The instability criterion for thermal modes is derived for longitudinal and transverse propagation in Section 4. Also numerical interpretation of the linear growth rate is done in Section 4. Finally, Section 5 contains the summary and discussion of the results.

## 2. Basic equations of the problem and perturbation

Let us consider an infinite homogeneous, radiating, thermally conducting, viscous plasma of finite electrical resistivity, including the effect of finite ion Larmor radius (FLR) corrections in the presence of magnetic field **B** (0, 0, B). The equations of the problem with these effects are

$$\frac{d\mathbf{v}}{dt} = -\frac{1}{\rho}\nabla p - \frac{\nabla\cdot\overleftrightarrow{\boldsymbol{P}}}{\rho} + \upsilon\nabla^2\mathbf{v} + \frac{1}{4\pi\rho}(\nabla\times\mathbf{B})\times\mathbf{B}\ , \tag{1}$$

$$\frac{d\rho}{dt} + \rho\nabla.\mathbf{v} = 0, \tag{2}$$

$$\frac{1}{\gamma-1}\frac{dp}{dt} - \frac{\gamma}{\gamma-1}\frac{p}{\rho}\frac{d\rho}{dt} + \rho L - \nabla.(\lambda\nabla T) = 0, \tag{3}$$

$$p = \rho RT, \tag{4}$$

$$\frac{\partial\mathbf{B}}{\partial t} = \nabla\times(\mathbf{v}\times\mathbf{B}) + \eta\nabla^2\mathbf{B}, \tag{5}$$

$$\nabla . \mathbf{B} = 0\,, \tag{6}$$

where $p$, $\rho$, $\upsilon$, $T$, $\eta$, $\lambda$, $R$, and $\gamma$ denote the fluid pressure, density, kinematic viscosity, temperature, electrical resistivity, thermal conductivity, gas constant and ratio of two specific heats (unperturbed) states, respectively. $L$ ($\rho$, $T$) is the radiative heat-loss function depends on local values of density and temperature of the fluid. The operator $(d/dt)$ is the Lagrangian derivative given as

$$(d/dt) = (\partial_t + \mathbf{v}.\nabla). \tag{7}$$

The components of pressure tensor $\overset{\leftrightarrow}{\boldsymbol{P}}$, considering the finite ion Larmor radius for the magnetic field along z-axis as given by Roberts & Taylor (1962) are

$$\begin{gathered}
P_{xx} = -\rho\upsilon_0\left(\frac{\partial \mathrm{v}_y}{\partial x} + \frac{\partial \mathrm{v}_x}{\partial y}\right), \qquad P_{yy} = \rho\upsilon_0\left(\frac{\partial \mathrm{v}_y}{\partial x} + \frac{\partial \mathrm{v}_x}{\partial y}\right), \\
P_{zz} = 0, \qquad P_{xy} = P_{yx} = \rho\upsilon_0\left(\frac{\partial \mathrm{v}_x}{\partial x} - \frac{\partial \mathrm{v}_y}{\partial y}\right), \\
P_{xz} = P_{zx} = -2\rho\upsilon_0\left(\frac{\partial \mathrm{v}_y}{\partial z} + \frac{\partial \mathrm{v}_z}{\partial y}\right), \qquad P_{yz} = P_{zy} = 2\rho\upsilon_0\left(\frac{\partial \mathrm{v}_z}{\partial x} + \frac{\partial \mathrm{v}_x}{\partial z}\right).
\end{gathered} \tag{8}$$

The parameter $\upsilon_0$ has the dimensions of the kinematics viscosity and defined as $\upsilon_0 = \Omega_L\, R_L^2/4$, where $R_L$ is the ion-Larmor radius and $\Omega_L$ is the ion gyration frequency.

The perturbation in fluid pressure, density, temperature, velocity, magnetic field, pressure tensor and heat-loss function is given as $\delta p$, $\delta\rho$, $\delta T$, $\mathbf{v}(\mathrm{v}_x, \mathrm{v}_y, \mathrm{v}_z)$, $\boldsymbol{\delta B}$( $\delta B_x$, $\delta B_y$, $\delta B_z$ ), $\boldsymbol{\delta}\overset{\leftrightarrow}{\boldsymbol{P}}$ and $L$ respectively. The perturbation state is given as

$$\begin{gathered}
p = p_0 + \delta p, \quad \rho = \rho_0 + \delta\rho, \quad T = T_0 + \delta T, \quad \mathbf{v} = \mathbf{v_0} + \mathbf{v} \text{ (with } \mathbf{v_0} = 0), \\
\boldsymbol{B} = \boldsymbol{B_0} + \boldsymbol{\delta B}, \quad \overset{\leftrightarrow}{\boldsymbol{P}} = \overset{\leftrightarrow}{\boldsymbol{P}}_0 + \boldsymbol{\delta}\overset{\leftrightarrow}{\boldsymbol{P}} \quad \text{and } L = L_0 + L \text{ (with } L_0 = 0),
\end{gathered} \tag{9}$$

where the suffix '0' represents the initial equilibrium state, which is independent of space and time. Substituting the perturbation state equation (9) into equations (1) to (7) and linearizing them by neglecting higher order perturbations, Equations (1)-(6) become

$$\partial_t \mathbf{v} = -\frac{1}{\rho}\nabla\delta p - \frac{\nabla\cdot\boldsymbol{\delta}\overset{\leftrightarrow}{\boldsymbol{P}}}{\rho} + \upsilon\nabla^2\mathbf{v} + \frac{1}{4\pi\rho}(\nabla\times\boldsymbol{\delta B})\times\boldsymbol{B}, \tag{10}$$

$$\partial_t \delta\rho + \rho \nabla.\mathbf{v} = 0, \tag{11}$$

$$\frac{1}{\gamma-1}\partial_t \delta p - \frac{\gamma}{\gamma-1}\frac{p}{\rho}\partial_t \delta\rho + \rho(L_\rho \delta\rho + L_T \delta T) - \lambda \nabla^2 \delta T = 0, \tag{12}$$

$$\frac{\delta p}{p} = \frac{\delta T}{T} + \frac{\delta\rho}{\rho}, \tag{13}$$

$$\partial_t \boldsymbol{\delta B} = \nabla \times (\mathbf{v} \times \boldsymbol{B}) + \eta \nabla^2 \boldsymbol{\delta B}, \tag{14}$$

$$\nabla.\boldsymbol{\delta B} = 0, \tag{15}$$

where $L_T$ and $L_\rho$ respectively denote partial derivatives $(\partial L / \partial T)_\rho$ and $(\partial L / \partial \rho)_T$ of the heat-loss function evaluated for the initial (unperturbed) state.

**3. Dispersion relation**

We seek plain wave solution of the form

$$\exp(ik_x x + i k_z z + i \sigma t), \tag{16}$$

where $\sigma$ is the frequency of harmonic disturbance, $k_x$ and $k_z$ are the wave numbers of the perturbations along x and z axes. Using equation (16) in linearized perturbation equations we get four equations and these four equations can be written in the following matrix form

$$\begin{bmatrix} P & F & 0 & \frac{ik_x}{k^2}\Omega_T^2 \\ -F & Q & -2\upsilon_0 k_x k_z & 0 \\ 0 & 2\upsilon_0 k_x k_z & M & \frac{ik_z}{k^2}\Omega_T^2 \\ ik_x \frac{\mathrm{V}^2 k^2}{d} & ik_x \upsilon_0 (k_x^2 + 4k_z^2) & 0 & -R \end{bmatrix} \begin{bmatrix} \mathrm{v}_x \\ \mathrm{v}_y \\ \mathrm{v}_z \\ s \end{bmatrix} = 0, \tag{17}$$

where

$$V^2=\frac{B^2}{4\pi\rho},\ \Omega_T^2=\frac{\Omega_I^2+\omega\Omega_j^2}{\omega+\beta},\ \Omega_j^2=c^2k^2,\ \ \Omega_I^2=k^2A,\ \ A=(\gamma-1)\left(TL_T-\rho L_\rho+\frac{\lambda k^2T}{\rho}\right),$$

$$\beta=(\gamma-1)\left(\frac{T\rho L_T}{p}+\frac{\lambda k^2T}{p}\right),\ \ L_T=(\partial L/\partial T)_\rho,L_\rho=(\partial L/\partial\rho)_T,\ M=\omega+\upsilon k^2,\ P=M+\frac{V^2k^2}{d},$$

$$Q=M+\frac{V^2k_z^2}{d},\ \ R=\omega^2+\omega\upsilon k^2+\Omega_T^2,\ \ F=\upsilon_0\left(k_x^2+2k_z^2\right),\ d=\left(\omega+\eta k^2\right),\ \ i\sigma=\omega. \tag{18}$$

Here $c=(\gamma p/\rho)^{1/2}$ is the adiabatic velocity of sound in the medium, $c'=(p/\rho)^{1/2}$ is the isothermal velocity of sound in the medium, $s=\delta\rho/\rho$ is the condensation of the medium.

Setting the determinant of coefficient matrix equal to zero, the general dispersion relation is obtained as

$$\left(\omega^2+\omega\upsilon k^2+\Omega_T^2\right)\left[\omega+\upsilon k^2+\frac{V^2k^2}{d}\right]\left\{\left(\omega+\upsilon k^2\right)\left[\omega+\upsilon k^2+\frac{V^2k_z^2}{d}\right]+4\upsilon_0^2k_x^2k_z^2\right\}-\frac{2\upsilon_0^2k_x^2k_z^2}{k^2}\Omega_T^2$$

$$\times\left[\omega+\upsilon k^2+\frac{V^2k^2}{d}\right]\left(k_x^2+4k_z^2\right)+\left(\omega+\upsilon k^2\right)\left[\omega^2+\omega\upsilon k^2+\Omega_T^2\right]\upsilon_0^2\left(k_x^2+2k_z^2\right)^2+2\upsilon_0^2k_x^2k_z^2$$

$$\times\left(k_x^2+2k_z^2\right)\frac{V^2}{d}\Omega_T^2-\frac{\upsilon_0^2k_x^2}{k^2}\Omega_T^2\left(k_x^2+2k_z^2\right)\left(\omega+\upsilon k^2\right)\left(k_x^2+4k_z^2\right)-\frac{V^2k_x^2}{d}\Omega_T^2\left(\omega+\upsilon k^2\right)\left[\omega+\upsilon k^2+\frac{V^2k_z^2}{d}\right]$$

$$-4\upsilon_0^2k_x^4k_z^2\frac{V^2}{d}\Omega_T^2=0. \tag{19}$$

The dispersion relation (19) represents the simultaneous inclusion of FLR corrections, radiative heat-loss function, thermal conductivity, viscosity, magnetic field and finite electrical conductivity on thermal instability of plasma. In absence of radiative heat-loss function the general dispersion relation (19) is identical to that of Vaghela & Chhajlani (1989) for non-gravitating, non-porous and non-permeability case. In absence of radiative heat-loss function, thermal conductivity, finite electrical resistivity and viscosity the general dispersion relation (19) is identical to Sharma (1974) for non-rotational and for non-gravitating case. In absence of FLR corrections and radiative heat-loss function the general dispersion relation (19) is identical to Shaikh et al. (2008) neglecting Hall current and neglecting gravitation in their case. In absence of FLR corrections and viscosity dispersion relation (19) is identical to Bora & Talwar (1993) neglecting Hall current, electron inertia, in their case. Also in absence of FLR corrections, viscosity and finite conductivity dispersion relation (19) reduces to that obtained by Field (1965).

Thus we have obtained the modified dispersion relation of thermal instability including the combined effect of FLR corrections, radiative heat-loss function, viscosity, thermal conductivity, finite electrical resistivity and magnetic field. Now we discuss the general dispersion relation (19) for longitudinal and transverse wave propagation.

## 4. Discussion

### 4.1. Longitudinal mode of propagation $(k_x = 0, \quad k_z = k)$

In this case the perturbations are taken parallel to the direction of the magnetic field $(i.e.\ k_x = 0, k_z = k)$. The dispersion relation (19) reduces to

$$\left(\omega+\upsilon k^2\right)\times\left\{\left[\omega+\upsilon k^2+\frac{V^2k^2}{\omega+\eta k^2}\right]^2+4\upsilon_0^2k^4\right\}\times\left[\omega^2+\omega\,\upsilon k^2+\frac{\Omega_I^2+\omega\Omega_j^2}{\omega+\beta}\right]=0. \quad (20)$$

The above dispersion relation shows the combined effect of FLR corrections, radiative heat-loss function, thermal conductivity, viscosity, magnetic field strength and finite electrical resistivity on thermal instability of plasma. On multiplying all the factors of equation (20) we get the dispersion relation, which is an eighth-order equation in $\omega$ and it is cumbersome to write such a lengthy equation. If we remove the effect of FLR corrections and viscosity in the above relation then we recover the relation given by Bora & Talwar (1993) excluding Hall current, electron inertia and self-gravitation in their case. Hence the above dispersion relation is the modified form of equation (21) of Bora & Talwar (1993) due to the inclusion of FLR corrections and viscosity in our case and by neglecting Hall current and electron inertia in their case for longitudinal propagation in dimensional form. But condition of instability is unaffected by the presence of FLR correction and viscosity. Thus we conclude that FLR corrections and viscosity have no effect on the condition of instability, but presence of these parameters modifies the growth rate of instability in the present case. Hence these are the new findings in our case than that of Bora & Talwar (1993). Also on comparing our dispersion relation (20) with dispersion relation (20) of Vaghela & Chhajlani (1989) we find that first two factors are the same but the third factor is different and gets modified because of radiative terms.

The dispersion relation (20) has three different factors and our aim is to take out the physics involved in each factor. So we discuss each factor separately. The first factor of the dispersion relation (20) gives

$$\omega+\upsilon k^2=0. \quad (21)$$

This represents a stable damped mode modified by the presence of viscosity of the medium. Thus viscous force is capable to stabilize the considered system. The above mode is unaffected by the presence of FLR correction, magnetic field strength, finite electrical resistivity, thermal conductivity and radiative heat-loss function.

The second factor of equation (20) gives

$$\omega^4 + 2\left(\upsilon k^2 + \eta k^2\right)\omega^3 + \left[\left(\upsilon k^2 + \eta k^2\right)^2 + 2\left(V^2 k^2 + \eta k^2 \upsilon k^2\right) + 4\upsilon_0^2 k^4\right]\omega^2 + \left[2\left(\upsilon k^2 + \eta k^2\right)\right.$$

$$\left.\times\left(V^2 k^2 + \eta k^2 \upsilon k^2\right) + 8\eta k^2 \upsilon_0^2 k^4\right]\omega + \left(V^2 k^2 + \eta k^2 \upsilon k^2\right)^2 + 4\eta^2 k^4 \upsilon_0^2 k^4 = 0. \tag{22}$$

The above dispersion relation represents the Alfven mode modified by the presence of FLR corrections, viscosity and finite electrical resistivity. But it is independent of thermal conductivity and radiative heat-loss function. Equation (22) is a fourth-order equation in $\omega$ having positive coefficients. According to Routh-Hurwitz's stability criterion, a necessary and sufficient condition of stability of the system is that the principal diagonal minors of Hurwitz matrix must be positive. On calculating we get all the principal diagonal minors positive. Hence equation (22) always represents stability.

The third factor of the dispersion relation (20) gives

$$\omega^3 + \left[\upsilon k^2 + (\gamma - 1)\left(\frac{T\rho L_T}{p} + \frac{\lambda k^2 T}{p}\right)\right]\omega^2 + \left[(\gamma - 1)\left(\frac{T\rho L_T}{p} + \frac{\lambda k^2 T}{p}\right)\upsilon k^2 + \left(c^2 k^2\right)\right]\omega$$

$$+\left\{k^2 (\gamma - 1)\left(T L_T - \rho L_\rho + \frac{\lambda k^2 T}{\rho}\right)\right\} = 0. \tag{23}$$

The above equation represents the combined influence of radiative heat-loss function, thermal conductivity and viscosity on the thermal instability of plasma. But there is no effect of FLR corrections, finite electrical conductivity and magnetic field on the thermal instability of the considered system. If the constant term of cubic equation (23) is less than zero this allows at least one positive real root which corresponds to the instability of the system. The condition of instability obtained from constant term of equation (23) is

$$\left\{k^2 (\gamma - 1)\left(T L_T - \rho L_\rho + \frac{\lambda k^2 T}{\rho}\right)\right\} < 0. \tag{24}$$

The above condition of instability is independent of FLR corrections, finite electrical conductivity, magnetic field strength and viscosity. The above inequality (24) is reduced form of Bora & Talwar (1993) neglecting self-gravitation in their case. Thus the dispersion relation and growth rate is modified due to FLR corrections and viscosity, but the condition of radiatively driven thermal instability is unaffected by the presence of FLR corrections and viscosity. The heat loss function is the prominent destabilizing process underlying the thermal instability that depends on temperature and density in the astrophysical objects like solar corona and star formation. The derivatives of heat-loss function decreases as temperature decrease and it increases as density increases i.e. $\left(L_T < 0\right)$ and $\left(L_\rho > 0\right)$. In this type of conditions a small temperature perturbation tends to grow naturally for example, on decreasing the temperature, cooling increases due to more radiative losses. In solar corona as the local temperature decreases, the local pressure decreases leading to the condensation of the cool plasma which radiates even faster because of an increase on density.

Equation (23) can be written in the form of Field (1965)

$$\omega^3 + c\left(\frac{\upsilon k^2}{c} + k_T + \frac{k^2}{k_\lambda}\right)\omega^2 + c^2\left[\frac{\upsilon k^2}{c} + \left(k_T + \frac{k^2}{k_\lambda}\right) + k^2\right]\omega + \frac{k^2 c^3}{\gamma}\left(k_T - k_\rho + \frac{k^2}{k_\lambda}\right) = 0, \tag{25}$$

where

$$k_T = \frac{(\gamma-1)L_T}{Rc}, \quad k_\lambda = \frac{Rc\rho}{(\gamma-1)\lambda}, \quad \text{and} \quad k_\rho = \frac{(\gamma-1)\rho L_\rho}{RcT}. \tag{26}$$

In dispersion relation (20) first and second factor give wave propagation, but third factor gives instability. In order to discuss the effect of each parameter on the growth rate of instability we solve equation (25) numerically by introducing the following dimensionless quantities.

$$\omega^* = \frac{\omega}{k_\rho c}, \quad \upsilon^* = \frac{\upsilon k_\rho}{c}, \quad k^* = \frac{k}{k_\rho}, \quad k_\lambda^* = \frac{k_\rho}{k_\lambda}, \quad k_T^* = \frac{k_T}{k_\rho}. \tag{27}$$

Using Eq. (27), we write Eq. (25) in non-dimensional form as

$$\omega^{*3} + \left[\upsilon^* k^{*2} + k_T^* + k_\lambda^* k^{*2}\right]\omega^{*2} + \left[\upsilon^* k^{*2}\left(k_T^* + k_\lambda^* k^{*2}\right) + k^{*2}\right]\omega^* + \frac{k^{*2}}{\gamma}\left(k_T^* - 1 + k_\lambda^* k^{*2}\right) = 0. \tag{28}$$

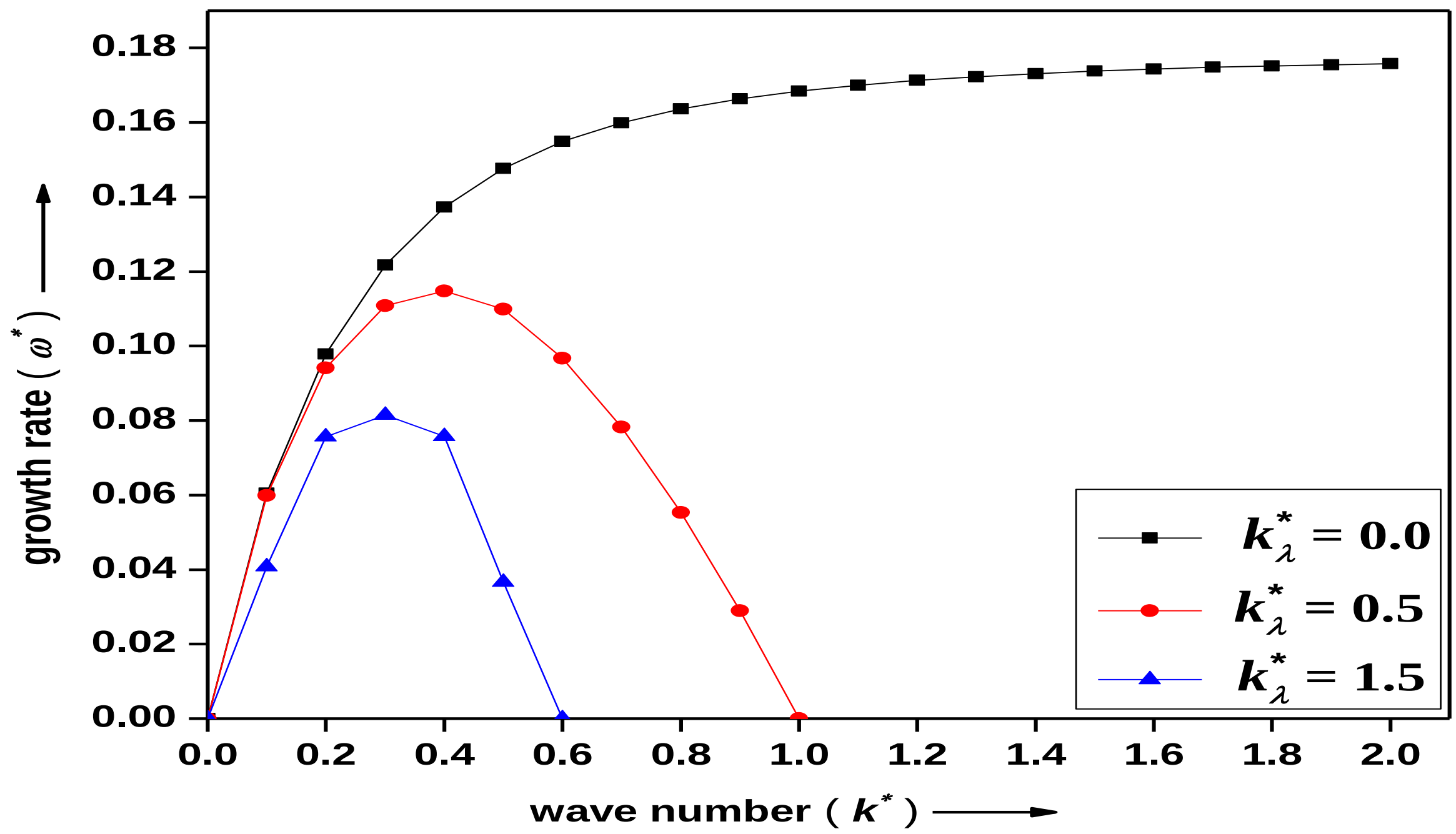

**Fig. 1:** The normalized growth rate ($\omega^*$) as a function of normalized wave number ($k^*$) for different values of $k_\lambda^*$ with $k_T^* = 0.5$and $v^* = 1$.

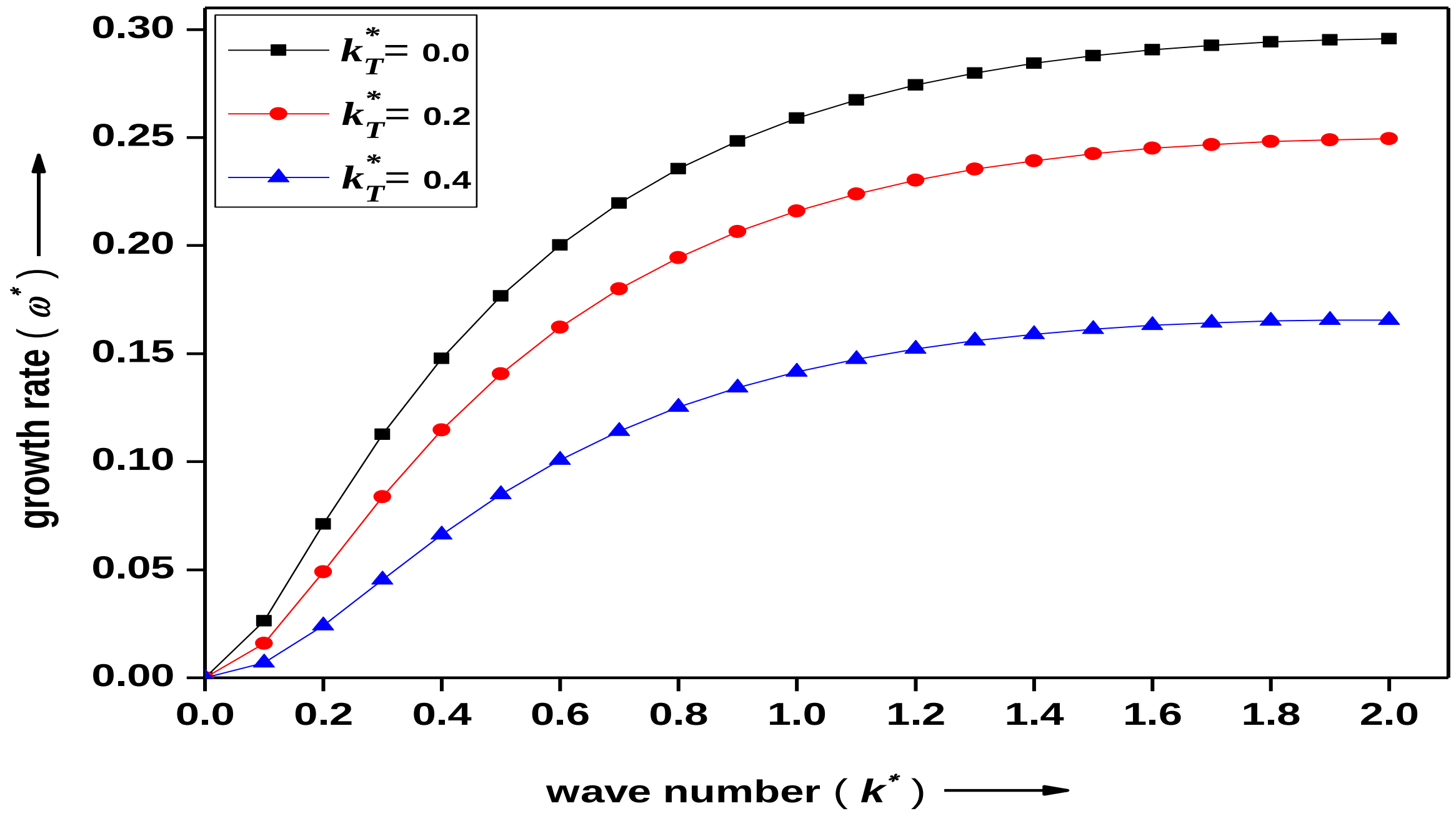

**Fig. 2:** The normalized growth rate ($\omega^*$) as a function of normalized wave number ($k^*$) for different values of $k_T^*$ with $k_\lambda^* = 0.01$ and $v^* = 1$.

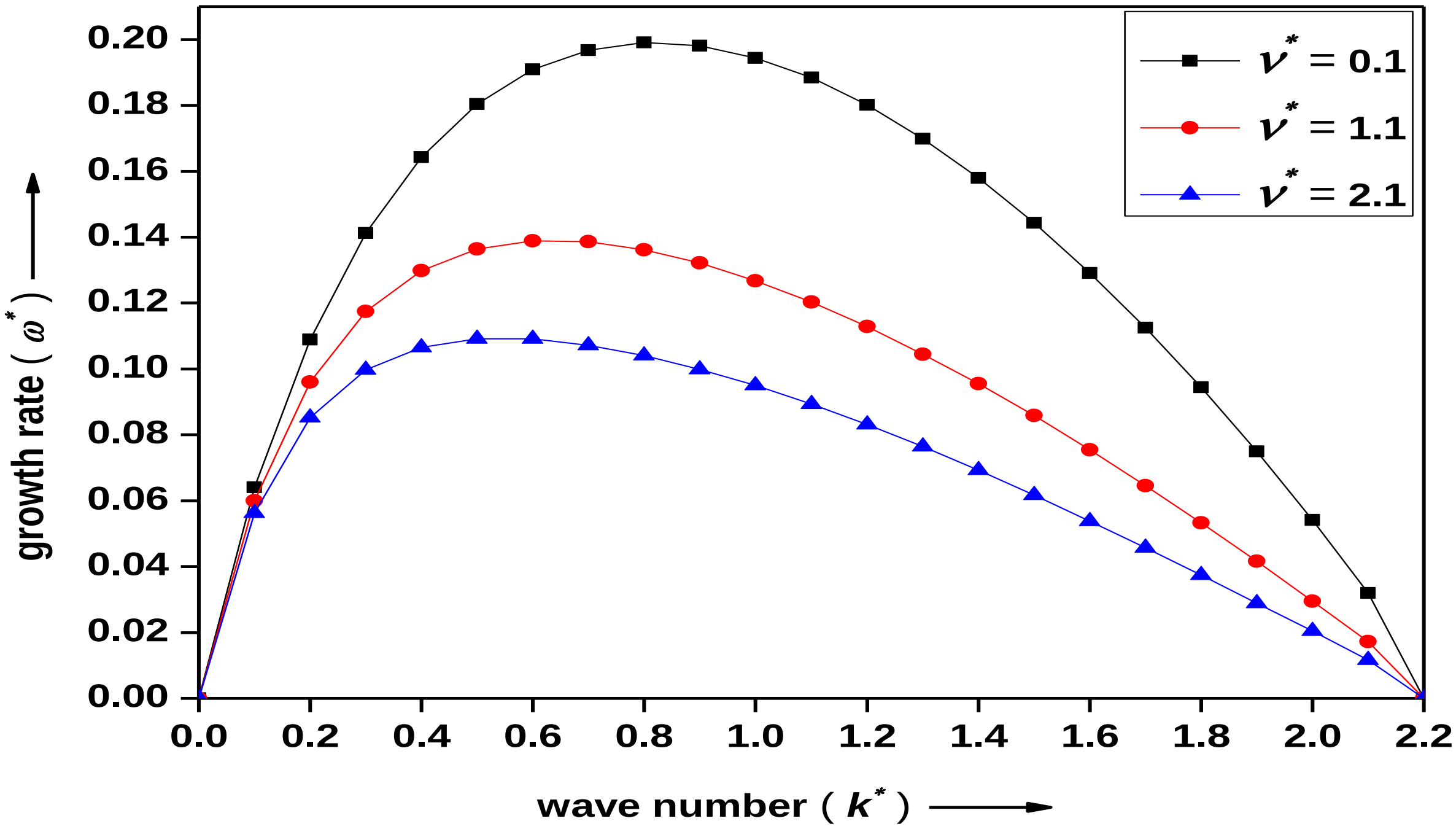

**Fig. 3:** The normalized growth rate ($\omega^*$) as a function of normalized wave number ($k^*$) for different values of $\nu^*$ with $k_\lambda^* = 0.1$ and $k_T^* = 0.5$

Figure 1 shows the effect of $k_\lambda^*$ on the growth rate of thermal instability for fixed values of other parameters. From curves it is clear that as the value of $k_\lambda^*$ increases both the peak value and the growth rate of thermal instability decreases. System becomes more stable with higher parameter $k_\lambda^*$. In Fig. 2 the plot of the growth rate of thermal instability against wave number for different values of the parameter $k_T^*$ is represented. From figure 2 it is conclude that as the value of $k_T^*$ increase, the peak value of curves decreases and the growth rate also decrease. Hence, the presence of $k_T^*$ also stabilizes the system. In Fig. 3 the effect of viscosity on the growth rate of thermal instability is represented. Figure 3 displays that on increasing the value of viscosity the growth rate of thermal instability decreases. Therefore, the parameters $k_\lambda^*$, $k_T^*$ and $\nu^*$ viscosity stabilize the system.

Thus we conclude that for longitudinal wave propagation as given by equation (20) the system is unstable only for modified thermal condition of instability, else it is stable. Also the modified thermal criterion remains unaffected by FLR corrections, viscosity, magnetic field and finite electrical resistivity, but radiative heat-loss function and thermal conductivity modify the expression and the fundamental thermal instability criterion becomes radiatively driven thermal instability criterion. From the curves we find that heat-loss function and viscosity have stabilizing influence.

### 4.2. Transverse mode of propagation $(k_x = k,\ k_z = 0)$

In this case the perturbations are taken perpendicular to the direction of the magnetic field $(i.e.\ k_x = k,\ k_z = 0)$. The dispersion relation (19) reduces to

$$\left(\omega+\upsilon k^2\right)^2\left\{\left(\omega+\upsilon k^2\right)\left[\omega^2+\omega\upsilon k^2+\frac{\omega V^2k^2}{\omega+\eta k^2}+\frac{\Omega_I^2+\omega\Omega_j^2}{\omega+\beta}\right]+\omega\upsilon_0^2k^4\right\}=0. \tag{30}$$

We find that for transverse mode of propagation the dispersion relation is modified due to the presence of FLR corrections, radiative heat-loss function, thermal conductivity, viscosity, finite electrical resistivity and magnetic field. The dispersion relation (30) has two different factors. The first factor of the dispersion relation (30) represents a damped mode modified by the presence of viscosity of the medium and is discussed in equation (21).

The second factor of the dispersion relation (30) gives

$$\omega^5+\left[2\upsilon k^2+\eta k^2+(\gamma-1)\left(\frac{T\rho L_T}{p}+\frac{\lambda k^2T}{p}\right)\right]\omega^4+\left\{2\upsilon k^2\left[\eta k^2+(\gamma-1)\left(\frac{T\rho L_T}{p}+\frac{\lambda k^2T}{p}\right)\right]\right.$$

$$\left.+\upsilon^2k^4+\eta k^2(\gamma-1)\left(\frac{T\rho L_T}{p}+\frac{\lambda k^2T}{p}\right)+V^2k^2+\upsilon_0^2k^4+c^2k^2\right\}\omega^3+\left\{2\eta k^2\upsilon k^2(\gamma-1)\right.$$

$$\times\left(\frac{T\rho L_T}{p}+\frac{\lambda k^2T}{p}\right)+\left[\eta k^2+(\gamma-1)\left(\frac{T\rho L_T}{p}+\frac{\lambda k^2T}{p}\right)\right]\left(\upsilon^2k^4+\upsilon_0^2k^4\right)+V^2k^2\left[\upsilon k^2+(\gamma-1)\right.$$

$$\left.\left.\times\left(\frac{T\rho L_T}{p}+\frac{\lambda k^2T}{p}\right)\right]+\left(c^2k^2\right)\left(\upsilon k^2+\eta k^2\right)+(\gamma-1)\left[k^2\left(TL_T-\rho L_\rho+\frac{\lambda k^2T}{\rho}\right)\right]\right\}\omega^2$$

$$+\left\{\upsilon k^2\left[\eta k^2\upsilon k^2(\gamma-1)\left(\frac{T\rho L_T}{p}+\frac{\lambda k^2T}{p}\right)+V^2k^2(\gamma-1)\left(\frac{T\rho L_T}{p}+\frac{\lambda k^2T}{p}\right)+\eta k^2\left(c^2k^2\right)\right]\right.$$

$$\left.+\eta k^2\upsilon_0^2k^4(\gamma-1)\left(\frac{T\rho L_T}{p}+\frac{\lambda k^2T}{p}\right)+(\gamma-1)\left[k^2\left(TL_T-\rho L_\rho+\frac{\lambda k^2T}{\rho}\right)\right]\left(\upsilon k^2+\eta k^2\right)\right\}\omega$$

$$+\eta k^2 \upsilon k^2 \left[ k^2 (\gamma - 1) \left( TL_T - \rho L_\rho + \frac{\lambda k^2 T}{\rho} \right) \right] = 0. \tag{31}$$

The above fifth degree equation in $\omega$ represents the combined influence of FLR corrections, radiative heat-loss function, thermal conductivity, finite electrical resistivity, viscosity and magnetic field on thermal instability of plasma. When constant term of equation (31) is less than zero this allows at least one positive real root which corresponds to the instability of the system. The condition of instability obtained from constant term of equation (31) is given as

$$\left[ k^2 (\gamma - 1) \left( TL_T - \rho L_\rho + \frac{\lambda k^2 T}{\rho} \right) \right] < 0. \tag{32}$$

The above condition of instability is independent of FLR correction, finite electrical resistivity, viscosity and magnetic field strength and is discussed in equation (24). Thus from equation (31) we find that the modified condition of thermal instability remains the same whether we consider the effect of FLR corrections or not, but due to the presence of FLR corrections the growth rate of radiatively driven thermal instability is modified. From equation (31) we see that if derivative of heat-loss function decrease with density, thermal instability does not arises, but when the heat-loss function increases with density $\left(L_\rho > 0\right)$ thermal instability occurs if $\lambda < \left(\rho^2 L_\rho - \rho TL_T\right) / \left(k^2 T\right)$ and for purely density dependent heat-loss function, thermal instability occurs if $\lambda < \left(\rho^2 L_\rho\right) / \left(k^2 T\right)$.

We solve equation (31) numerically by introducing the following dimensionless quantities

$$\omega^* = \frac{\omega}{k_\rho c}, \quad \upsilon^* = \frac{\upsilon k_\rho}{c}, \quad k^* = \frac{k}{k_\rho}, \quad k_\lambda^* = \frac{k_\rho}{k_\lambda}, \quad k_T^* = \frac{k_T}{k_\rho}, \quad \eta^* = \frac{\eta k_\rho}{c}, \quad \upsilon_0{}^* = \frac{\upsilon_0 k_\rho}{c}. \tag{33}$$

Using Eq. (33), we write Eq. (31) in non-dimensional form as

$$\omega^{*5} + \left\{ 2\upsilon^* k^{*2} + \eta^* k^{*2} + k_T^* + k_\lambda^* k^{*2} \right\} \omega^{*4} + \left\{ 2\upsilon^* k^{*2} \left[ k_T^* + k_\lambda^* k^{*2} + \eta^* k^{*2} \right] + \upsilon^{*2} k^{*4} + \eta^* k^{*2} \left[ k_T^* + k_\lambda^* k^{*2} \right] \right.$$

$$\left. + V^{*2} k^{*2} + \upsilon_0^{*2} k^{*4} + k^{*2} \right\} \omega^{*3} + \left\{ 2\eta^* k^{*2} \upsilon^* k^{*2} \left[ k_T^* + k_\lambda^* k^{*2} \right] + \left[ \eta^* k^{*2} + k_T^* + k_\lambda^* k^{*2} \right] + V^{*2} k^{*2} \right.$$

$$\left. \times \upsilon^* k^{*2} + k_T^* + k_\lambda^* k^{*2} \right] + k^{*2} \left[ \upsilon^* k^{*2} + \eta^* k^{*2} \right] + \frac{k^{*2}}{\gamma} \left[ k_T^* + k_\lambda^* k^{*2} - 1 \right] \right\} \omega^{*2} + \upsilon_0^* k^{*2} \eta^* k^{*2} \left[ k_T^* + k_\lambda^* k^{*2} \right] + \frac{k^{*2}}{\gamma} \left[ k_T^* + k_\lambda^* k^{*2} - 1 \right]$$

$$\times\left(\upsilon^* k^{*2}+\eta k^{*2}\right)\Big\}\omega^*+\upsilon^* k^{*2}\eta k^{*2}\left\{\frac{k^{*2}}{\gamma}\left[k_T^*+k_\lambda^* k^{*2}-1\right]\right\}=0. \tag{34}$$

In Figs. 4-8 the dimensionless growth rate ($\omega^*$) has been plotted against the dimensionless wave number ($k^*$) to see the effect of various physical parameters such as radiative heat-loss function, resistivity and FLR corrections on the growth rate of thermal instability. From Fig. 4 it is clear that as the value of $k_\lambda^*$ increases the growth rate decreases. Thus the parameter $k_\lambda^*$ has a stabilizing effect on the system. From Fig. 5 it is conclude that growth rate decreases with increasing parameter $k_T^*$. Thus the presence of $k_T^*$ stabilize the system. Figure 6 displays the influence of viscosity on the thermal instability. From figure 6 it is clear that the viscosity has a stabilizing effect on the thermal instability. Figure 7 shows the effect of resistivity on the thermal instability. From curves it is clear that the growth rate of thermal instability increases as the value of resistivity increases. Hence the resistivity has destabilizing influence on the system. Figure 8 displays the influence of FLR corrections on the thermal instability. From figure 8 it is clear that the FLR corrections stabilize the radiative driven thermal instability. Therefore, the parameters radiative heat-loss functions FLR corrections and viscosity have stabilizing influence on the system while the resistivity has destabilizing influence on the system.

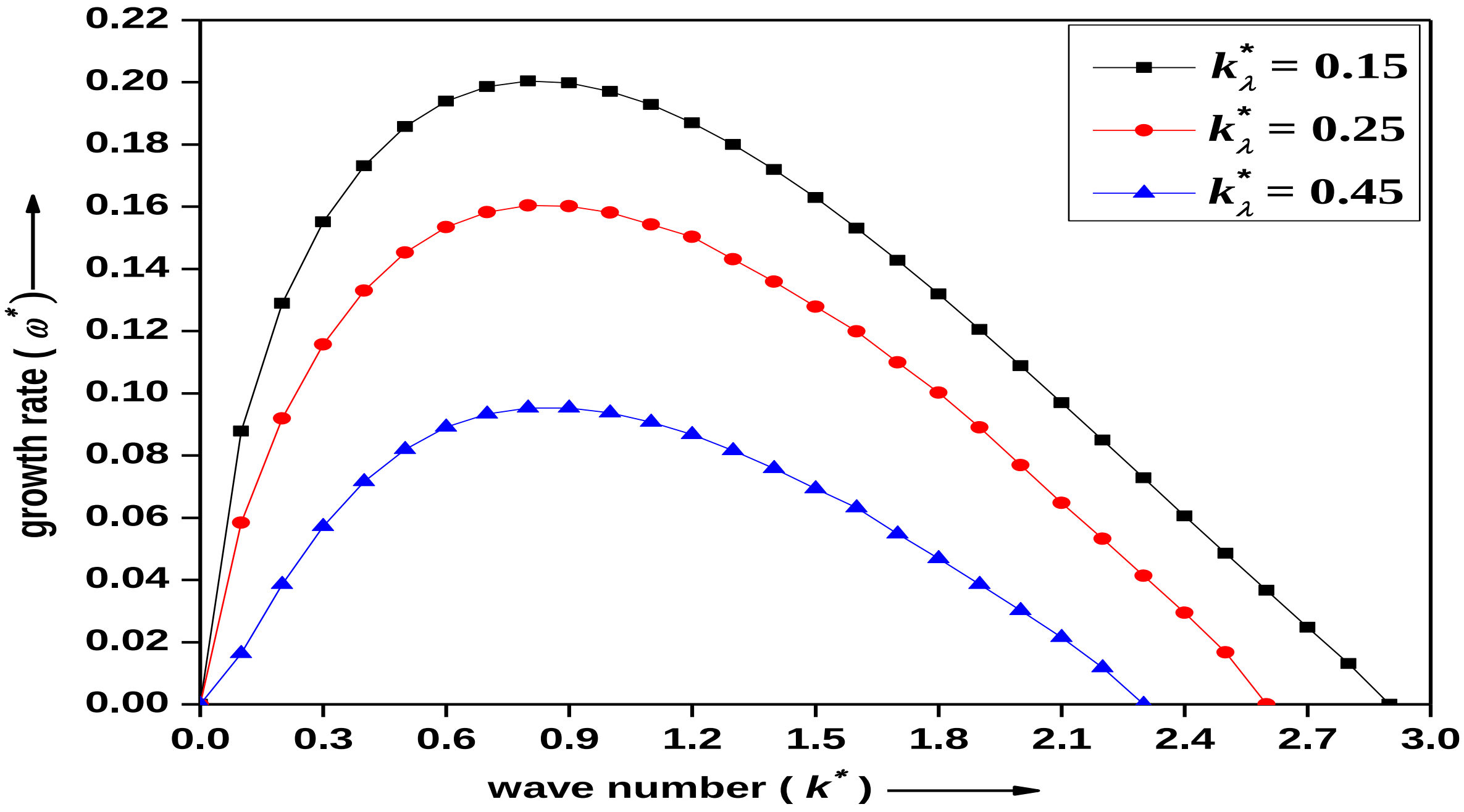

**Fig. 4:** The normalized growth rate ($\omega^*$) as a function of normalized wave number ($k^*$) for different values of $k_\lambda^*$ having $\eta^*=v^*=v_0^*=1$, and $k_T^*=0.5$.

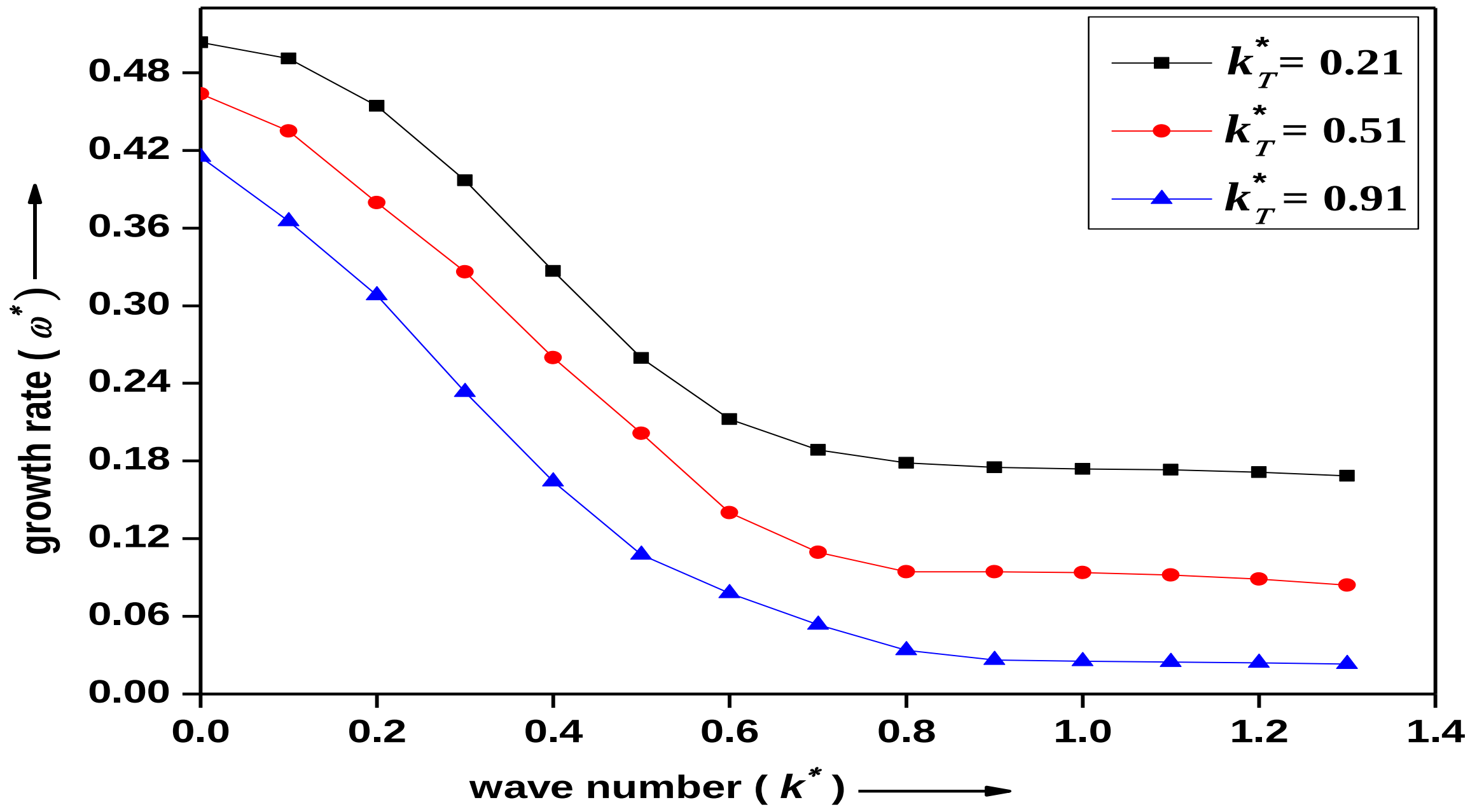

**Fig. 5:** The normalized growth rate ($\omega^*$) as a function of normalized wave number ($k^*$) for different values of $k_T^*$ having $\eta^* = v^* = v_0^* = 1$, and $k_\lambda^* = 0.1$.

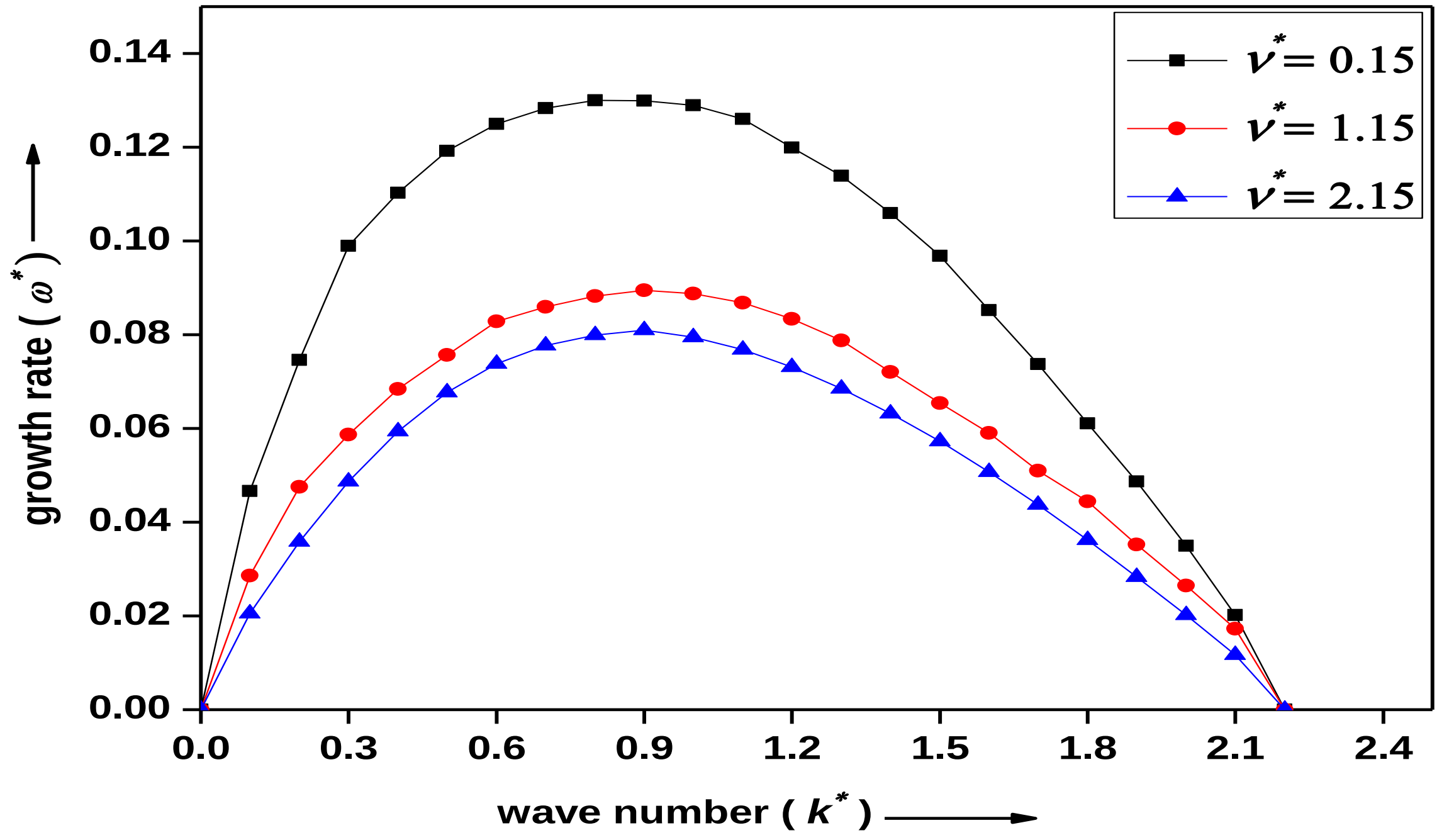

**Fig. 6:** The normalized growth rate ($\omega^*$) as a function of normalized wave number ($k^*$) for different values of $v^*$ having $\eta^* = v_0^* = 1$, with $k_T^* = 0.5$ and $k_\lambda^* = 0.1$.

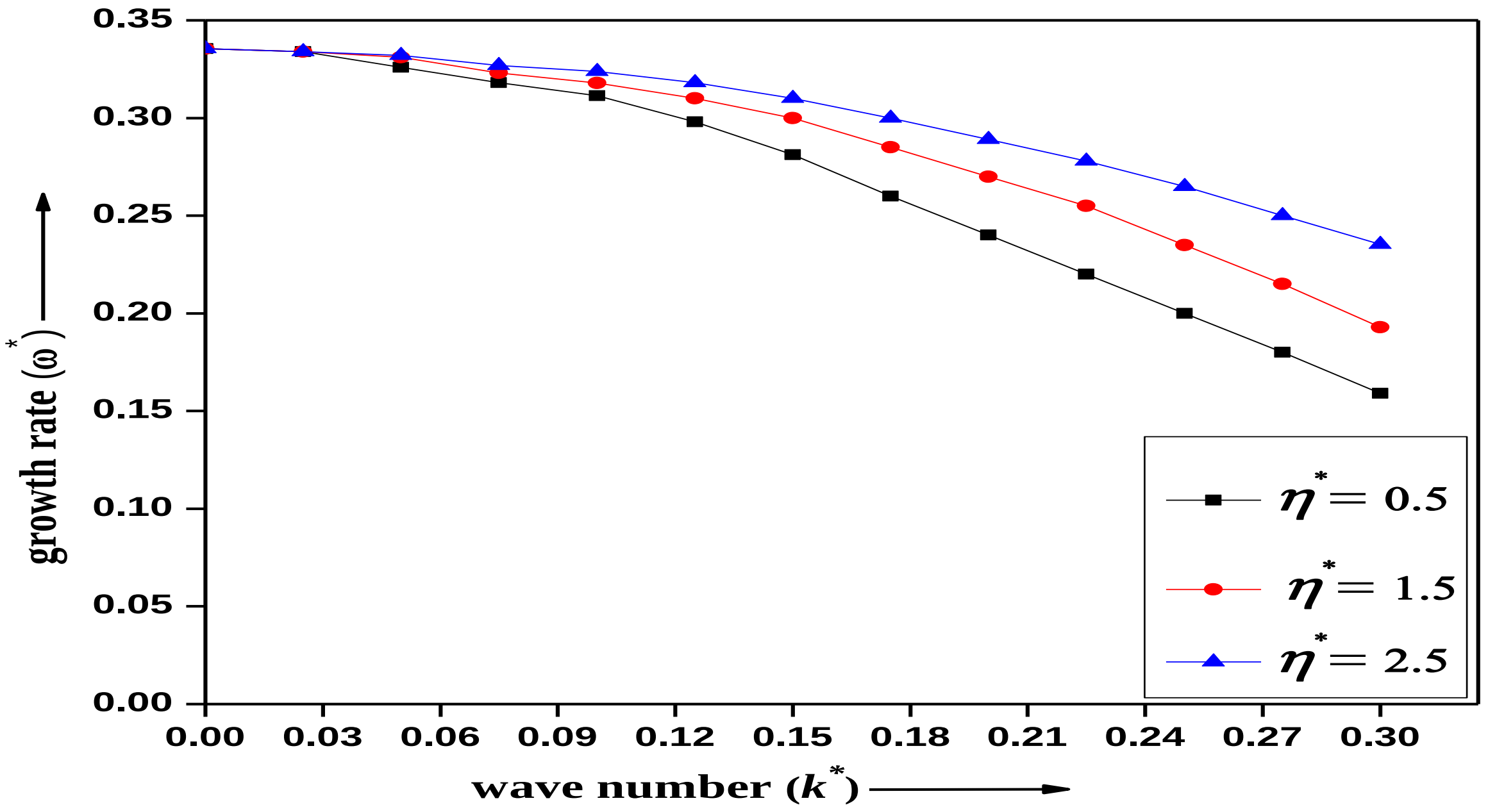

**Fig. 7:** The normalized growth rate ($\omega^*$) as a function of normalized wave number ($k^*$) for different values of $\eta^*$, having $k_T^* = 0.5$ with $v^* = v_0^* = 1$ and $k_\lambda^* = 0.1$.

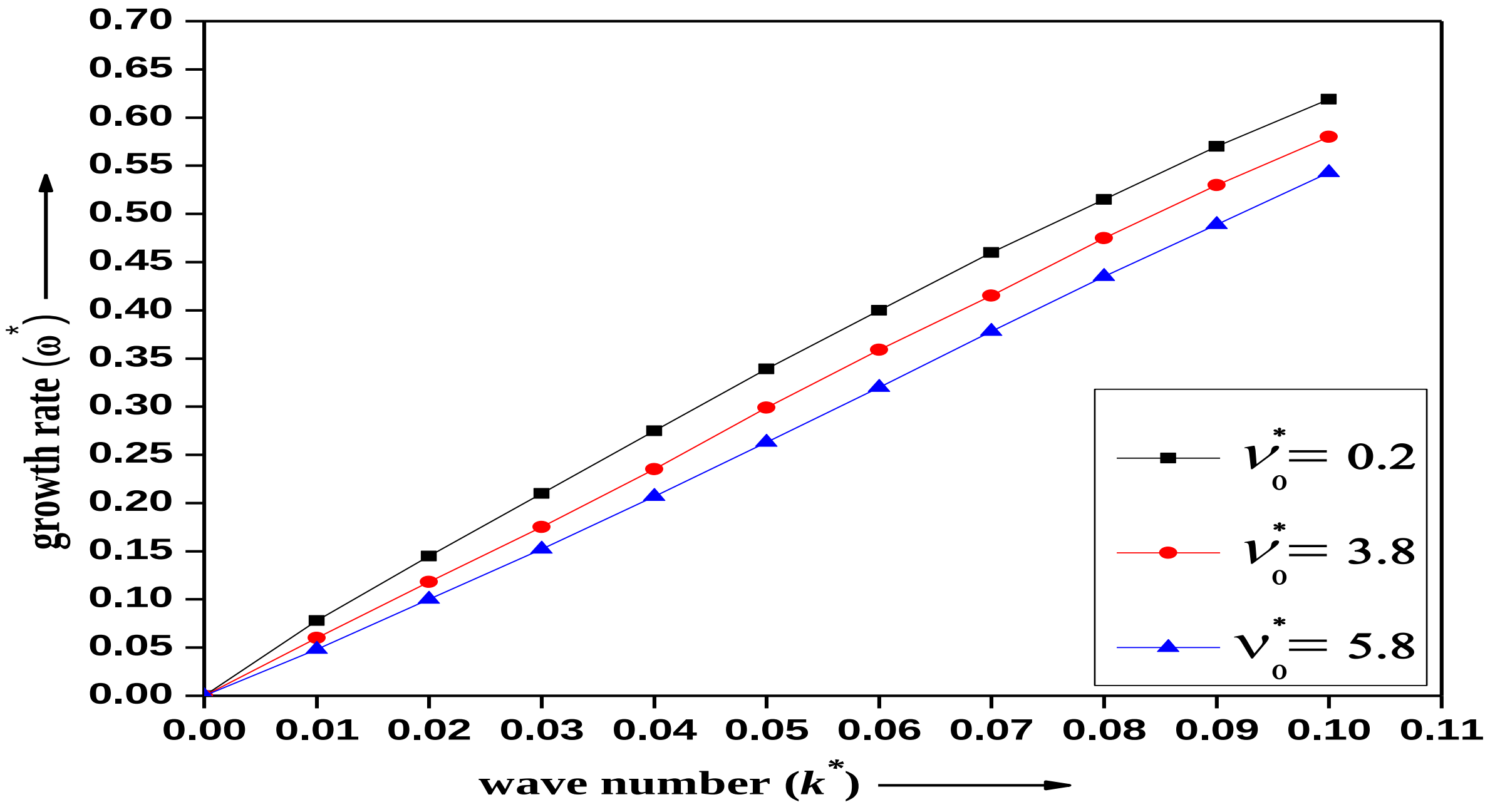

**Fig. 8:** The normalized growth rate ($\omega^*$) as a function of normalized wave number ($k^*$) for different values of $v_0^*$, having $k_T^* = 0.3$ with $v^* = \eta^* = 1$ and $k_\lambda^* = 0.2$.

Now we wish to examine the effect of FLR corrections and radiative heat-loss function on the considered system with some simplifications such as viscous and non-viscous medium, and at the same time we wish to investigate the physics involved in such simplifications in the present problem.

**4.2.2. For non-viscous medium**

**4.2.2.1. Effect of FLR corrections**

In absence of viscosity, but in presence of FLR corrections $(\upsilon = 0,$ $\eta = L_{T,\rho} = V = \lambda = \upsilon_0 \neq 0)$ equation (31) becomes

$$\omega^4 + \left[\eta k^2 + (\gamma-1)\left(\frac{T\rho L_T}{p} + \frac{\lambda k^2 T}{p}\right)\right]\omega^3 + \left[\eta k^2(\gamma-1)\left(\frac{T\rho L_T}{p} + \frac{\lambda k^2 T}{p}\right) + V^2k^2 + \upsilon_0^2 k^4 + c^2k^2\right]\omega^2$$

$$+\left\{\upsilon_0^2 k^4\left[\eta k^2 + (\gamma-1)\left(\frac{T\rho L_T}{p} + \frac{\lambda k^2 T}{p}\right)\right] + V^2k^2(\gamma-1)\left(\frac{T\rho L_T}{p} + \frac{\lambda k^2 T}{p}\right) + \eta k^2\left(c^2k^2\right)\right.$$

$$\left.+\left[k^2(\gamma-1)\left(TL_T - \rho L_\rho + \frac{\lambda k^2 T}{\rho}\right)\right]\right\}\omega + \eta k^2\left\{\upsilon_0^2 k^4(\gamma-1)\left(\frac{T\rho L_T}{p} + \frac{\lambda k^2 T}{p}\right)\right.$$

$$\left.+k^2(\gamma-1)\left(TL_T - \rho L_\rho + \frac{\lambda k^2 T}{\rho}\right)\right\} = 0. \tag{35}$$

The condition of instability obtained from constant term of equation (35) is given as

$$\left\{\upsilon_0^2 k^4(\gamma-1)\left(\frac{T\rho L_T}{p} + \frac{\lambda k^2 T}{p}\right) + k^2(\gamma-1)\left(TL_T - \rho L_\rho + \frac{\lambda k^2 T}{\rho}\right)\right\} < 0. \tag{36}$$

From the above condition of instability given by equation (36) it is conclude that FLR correction tries to stabilize the radiatively driven thermal instability. Also on comparing equations (31) and (35) we see that inclusion of viscosity removes the effects of FLR correction from condition of instability. So in both the cases either the system is viscous or non-viscous, FLR correction has a stabilizing effect on the system. Thus in presence of FLR correction the condition of instability given by Bora & Talwar (1993) is modified as given by equation (36). In absence of FLR corrections equation (35) is identical to Bora & Talwar (1993), thus the dispersion relation is modified due to the inclusion of effects of FLR in the present system. Thus we find that due to the presence of FLR corrections criterion of thermal instability is modified and also the growth rate of instability in this case is modified with the inclusion of FLR correction in the present MHD set of equations. From equation (36) we see that if a heat-loss function decrease with density, thermal instability does not arises, but when the heat-loss function increases with density $\left(L_\rho > 0\right)$ thermal instability occurs if $\lambda < \dfrac{\left(\rho L_\rho - T L_T\right)}{\upsilon_0^2 \rho L_T \left(\upsilon_0^2 k^4 + k^2\right)}$ and for purely density dependent heat-loss function, thermal instability occurs if $\lambda < \dfrac{\rho L_\rho}{k^2 T\left(\upsilon_0^2 k^4 + k^2\right)}$.

**4.2.2.3. Effect of electrical conductivity with FLR corrections**

In absence of viscosity and finite electrical resistivity, but in presence of FLR ($\upsilon = \eta = 0$, $L_{T,\rho} = V = \lambda = \upsilon_0 \neq 0$) the condition of instability obtained from equation (31) is given as

$$\left\{\left(\upsilon_0^2 k^4 + V^2 k^2\right)(\gamma - 1)\left(\frac{T\rho L_T}{p} + \frac{\lambda k^2 T}{p}\right) + k^2(\gamma - 1)\left(T L_T - \rho L_\rho + \frac{\lambda k^2 T}{\rho}\right)\right\} < 0. \tag{37}$$

From the above condition of instability given by equation (37) we conclude that FLR correction and magnetic field try to stabilize the radiatively driven thermal instability. So in both the cases whether the system is finitely conducting or infinitely conducting magnetic field and FLR correction tries to stabilize the radiatively driven thermal instability. It is clear that if a heat-loss function decrease with density, thermal instability does not arises, but when the heat-loss function increases with density $\left(L_\rho > 0\right)$ thermal instability occurs if $\lambda < \dfrac{\left(\rho L_\rho - T L_T\right)}{\upsilon_0^2 \rho L_T \left(\upsilon_0^2 k^4 + V^2 k^2 + k^2\right)}$ and for purely density dependent heat-loss function, thermal instability occurs if $\lambda < \dfrac{\rho L_\rho}{k^2 T\left(\upsilon_0^2 k^4 + V^2 k^2 + k^2\right)}$.

In absence of FLR corrections the above condition of instability is identical to its Bora & Talwar (1993) excluding electron inertia and self-gravitation in that case. It is conclude that condition of instability given by Bora & Talwar (1993) is modified by inclusion of FLR corrections in our case. Thus the present results are the improvement of Bora & Talwar's ones (1993). From equation (31) it is clear that the dispersion relation given by Field (1965) is modified due to the presence of FLR correction, viscosity, magnetic field and finite electrical conductivity in our present case.

Thus it is clear that for transverse wave propagation the thermal instability criterion is affected by FLR corrections, radiative heat-loss functions, thermal conductivity, viscosity, magnetic field strength and finite electrical resistivity. From curves it is clear that FLR corrections, heat-loss function and viscosity have stabilizing influence, where as finite resistivity has destabilizing influence on the thermal instability of plasma.

## 5. Conclusions

Thus the study of thermal instability of an infinite homogeneous viscous electrically and thermally conducting fluid including the effects of FLR corrections and radiative heat-loss function for molecular cloud formation and star formation has been done. The general dispersion relation is obtained. It is modified due to the presence of considered physical parameters. Also dispersion relations are obtained for longitudinal and transverse mode of propagation to the direction of the magnetic field. It is found that thermal instability criterion remains valid and gets modified because of FLR corrections, radiative heat-loss function, thermal conductivity and magnetic field. The presence of thermal conductivity and radiative heat-loss function modify the fundamental criterion of thermal instability into radiatively driven thermal instability criterion. The viscosity parameter has a stabilizing effect on the system for both the longitudinal and transverse mode of propagation. For longitudinal wave propagation magnetic field, viscosity, finite electrical resistivity and FLR correction have no effect on thermal instability criterion, but FLR corrections and finite electrical resistivity modify the growth rate of Alfven mode.

For transverse wave propagation FLR corrections stabilize the system in all the cases, but it modifies the condition of instability only for the case of non-viscous medium. Also, magnetic field stabilizes the system. Numerical calculations have shown that the heat-loss function, FLR corrections and viscosity have stabilizing effect, whereas finite resistivity has a destabilizing effect on the thermal instability.

In this paper the linear analysis of thermal instability in a collapsing warm gas cloud is performed considering the effect of different parameters. When the cloud density reaches critical value, the cloud fragments into cool dense condensations via thermal instability. The critical density increases as metallicity decreases. Condensations collide with each other, and self-gravitating clumps are produced when the mean cloud density becomes adequately high; then stars form. Expansion of the H II region around the massive star and supernova explosions will blow off neighboring gas and discontinue the star formation process. When the mean density is high during star formation, the high virial velocity prevents to the expansion of the H II region. Also, in such high-density environments, the star formation timescale is smaller than the lifetime of a massive star. Then the gas in the cluster-forming region is converted into stars efficiently, before the gas is diffused by expanding H II region or supernova explosion. Thus, it is suggested that high star formation efficiency and bound cluster formation are expected in low-metallicity and strong-radiation environments. Such environments exist in dwarf galaxies, the early stage of our galaxy, and starburst galaxies (Hobbs et al. 2013).

**Acknowledgements**

Author (SK) is grateful to Engineer Praveen Vashistha, Chairman Mahakal Institute of Technology, and Dr. Vivek Bansod, Director MIT Ujjain, for continuous support.